\documentclass[useAMS,usenatbib]{mn2e}
\usepackage[dvips]{epsfig}
\def \be{\begin{equation}}
\def \ee{\end{equation}}
\def \bdm{\begin{eqnarray}}
\def \edm{\end{eqnarray}}
\title[Nonlinear damping and CR transport]{Nonlinear damping of slab modes and cosmic ray transport}
\author[A. Shalchi et al.]{A. Shalchi$^{1}$, A. Lazarian$^{2}$, and R. Schlickeiser$^{1}$ \\
$^{1}$Institut f\"ur Theoretische Physik, Lehrstuhl IV: Weltraum- und Astrophysik, Ruhr-Universit\"at Bochum, D-44780 Bochum, Germany\\
$^{2}$Astronomy Department, University of Wisconsin, Madison, Wisconsin 53706, USA}
\begin{document}
\date{Accepted / Received}
\maketitle
\begin{abstract}
By applying recent results for the slab correlation time scale onto cosmic ray scattering theory, we compute
cosmic ray parallel mean free paths within the quasilinear limit. By employing these results onto charged
particle transport in the solar system, we demonstrate that much larger parallel mean free paths
can be obtained in comparison to previous results. A comparison with solar wind observations is also
presented to show that the new theoretical results are much closer to the observations than the previous
results.
\end{abstract}
\section{Introduction}
Cosmic rays (CRs) interacting with turbulent magnetic fields get scattered and accelerated (see Melrose 1968,
Schlickeiser 2002). The theoretical description of these scattering and acceleration processes are essential
for understanding the penetration and modulation of low-energy cosmic rays in the heliosphere, the confinement 
and escape of galactic cosmic rays from the Galaxy, and the efficiency of diffusive shock acceleration mechanisms. 

A key factor in CR scattering are the properties of the magnetic fields. A standard approach
is the assumption of a superposition of  a mean magnetic field $\vec{B}_0 = B_0 \vec{e}_z$ and a turbulent component
$\delta \vec{B} (\vec{x})$. Whereas the mean field can easily be meassured in the solar system (here we find approximatelly
$B_0 \approx 4-5 nT$), the turbulent component has to be emulated by turbulence models. In the literature there
is no consensus available about the true turbulence properties (see Cho \& Lazarian 2005 for a review).
In the solar system, however, some turbulence properties such as the wave spectrum can be obtained from meassurements 
(see e.g. Denskat \& Neubauer 1983, Bruno \& Carbone 2005).

More unclear are the orientation of the turbulence wave vectors (also refered to as turbulence geometry) and the
dynamical decorrelation of the magnetic fields. In a recent CR diffusion study (Shalchi et al. 2006) a slab/2D
composite model was combined with a nonlinear anisotropic dynamical turbulence (NADT) model. This model
can be used to reproduce meassured CR mean free paths parallel and perpendicular to the mean field $\vec{B}_0$.
The authors of this article assumed that the slab correlation time scale is independent of the wave vector $\vec{k}$.

In a recent study (Lazarian \& Beresnyak 2006), however, it was shown that the slab time scale is indeed
$\vec{k}-$dependent.\footnote{That study put to the test the idea of the damping of slab perturbations by the
ambient turbulence in Yan \& Lazarian (2002), Farmer \& Goldreich (2004).}
More precisely it was found that $t_c^{-1} = \gamma_c = v_A \sqrt{k_{\parallel} / L}$; here
we used the correlation time $t_c$, the correlation rate $\gamma_c$, the Alfv\'en speed $v_A$, and the outer
scale of the turbulence $L$. It is the purpose of this article to apply this new result of the slab correlation 
time scale onto cosmic ray parallel diffusion. A comparison with solar wind observations of the parallel mean
free path is also presented. It is demonstrated that we can find a much larger parallel mean free path if
we employ the correlation time scale of Lazarian \& Beresnyak (2006).

In Section 2 we explain the turbulence model that is used in this article. In Section 3
a quasilinear description of cosmic ray scattering is combined with this turbulence model
to derive analytic forms of the pitch-angle diffusion coefficient and the parallel mean free path.
In Section 4 we evaluate these formulas numerically to compute diffusion coefficients and
we also provide a comparison with previous results and solar wind observations. In the
closing Section 5 our results are summerized.
\section{The turbulence model}
The key input into a cosmic ray transport theory like QLT is the tensor $P_{lm}$ which describes the correlation of the
turbulent magnetic fields:
\be
P_{lm} (\vec{k},t) = \left< \delta B_l (\vec{k},t) \delta B_m^{*} (\vec{k},0) \right>
\ee
Therefore the $\vec{k}$-dependence and the time-dependence of the correlation tensor $P_{lm} (\vec{k},t)$ have to be 
specified which is done in the next two subsections.
\subsection{The slab model for the turbulence geometry}
For mathematical simplicity we employ the often used slab model for the turbulence geometry. 
Physical processes that can induce slab modes are differential damping of fast mode
waves (Yan \& Lazarian 2002) or instabilities (Lazarian \& Beresnyak 2006).

By assuming the same temporal behavior of all tensor components we have
\be
P_{lm} (\vec{k},t) = P_{lm}^{slab} (\vec{k}) \cdot \Gamma^{slab} (k_{\parallel},t)
\label{c2e2}
\ee
with the dynamical correlation function $\Gamma^{slab} (k_{\parallel},t)$ and the time independent correlation tensor
$P_{lm}^{slab} (\vec{k})$. The tensor $P_{lm}^{slab} (\vec{k})$ is determined by the turbulence geometry and the wave spectrum 
whereas the function $\Gamma^{slab} (\vec{k},t)$ describes dynamical effects. 

For pure slab turbulence we have $\delta B_i (\vec{x}) = \delta B_i (z)$ and therefore
\bdm
P_{lm}^{slab} (\vec{k}) & = & g^{slab}(k_{\parallel}) {\delta (k_{\perp}) \over k_{\perp}} \nonumber\\
& \times & \left\{
\begin{array}{ccc}
\delta_{lm} - {k_{l} k_{m} \over k^2} & \textnormal{if} & \textnormal{l,m=x,y} \\
0 & \textnormal{if} & \textnormal{l or m = z}
\end{array}
\right.
\label{c2e3}
\edm
In Eq. (\ref{c2e3}) we assumed the case of vanishing magnetic helicity. The function $g^{slab} (k_{\parallel})$ is the slab wave spectrum
which can be approximated by a power-law spectrum with energy, inertial, and dissipation range (see Fig. 1):
\bdm
& & g^{slab} (k_{\parallel}) = {C(\nu) \over 2 \pi} l_{slab} \delta B_{slab}^2 \nonumber\\
& \times & \left\{
\begin{array}{ccc}
0 & \textnormal{for} & 0 \leq k_{\parallel} < k_{min} \\
(1 + k_{\parallel}^2 l_{slab}^2)^{-\nu} & \textnormal{for} & k_{min} \leq k_{\parallel} < k_{d} \\
(1+k_{d}^2 l_{slab}^2)^{-\nu} (k_{d} / k_{\parallel})^{p} & \textnormal{for} & k_{d} \leq k_{\parallel} \\
\end{array}
\right.
\label{c2e5}
\edm
where we used the normalization function
\be
C(\nu) = {1 \over 2 \sqrt{\pi}} {\Gamma (\nu) \over \Gamma (\nu - 1/2)}.
\ee
\begin{figure}
\begin{center}
\epsfig{file=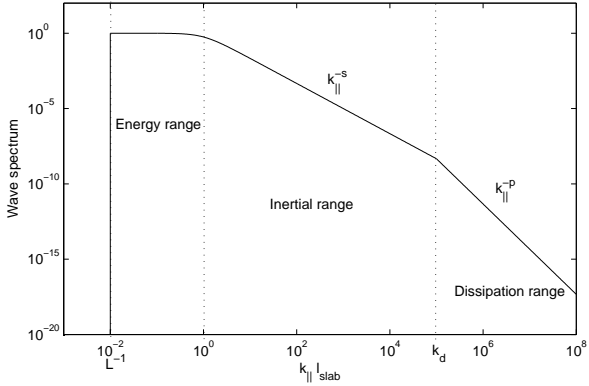, width=250pt}
\end{center}
\caption{The power spectrum used for our calculations. We used a general spectrum with energy, inertial, and dissipation range. 
The dissipation wavenumber $k_d$ divides the inertial range from the dissipation range.}
\label{nadtf1}
\end{figure}
Furthermore, we used the slab bendover scale $l_{slab}$, the dissipation wavenumber $k_{d}$, the turbulence strength 
$\delta B_{slab}^2$, the inertial range spectral index $s=2\nu$ and the dissipation range spectral index $p$. The parameter
$k_{min}=L^{-1}$ indicates the smallest wavenumber. Also the dynamical correlation function $\Gamma^{slab} (k_{\parallel},t)$ 
has to be determined which is done in the next subsection.
\subsection{Improved form of the slab correlation time scale}
In the past several models have been developed to approximate the dynamical correlation function $\Gamma^{slab} (k_{\parallel},t)$
(e.g. Schlickeiser \& Achatz 1993, Bieber et al. 1994, Shalchi et al. 2006). Some examples are given in 
Table \ref{tab1}.
\begin{table}
\caption{Previous models for the slab dynamical correlation function. Here $v_A$ is the Alfv\'en velocity, $\alpha$ is a 
parameter which allows to adjust the strength of dynamical effects, and $\beta = \alpha Z / l_{2D}$ is used in
the NADT (nonlinear anisotropic dynamical turbulence)-model 
($Z = \sqrt{2} \delta B_{2D} / \sqrt{4 \pi \rho_d} = \sqrt{2} v_A \delta B_{2D} / B_0$).}
\begin{tabular}{|l|l|}\hline
\vphantom{$1 \over 2$} $ \textnormal{Model}									$ & $ \Gamma ( \vec{k}, t)						$ \\ 
\hline\hline
\vphantom{$1 \over 2$} $ \textnormal{Magnetostatic model}						$ & $ 1										$ \\
\vphantom{$1 \over 2$} $ \textnormal{Damping model of dynamical turbulence}		$ & $ e^{-\alpha v_A \mid k \mid t}				$ \\
\vphantom{$1 \over 2$} $ \textnormal{Random sweeping model}					$ & $ e^{-(\alpha v_A k t)^2}					$ \\
\vphantom{$1 \over 2$} $ \textnormal{Plasma wave model for shear Alfv\'en waves}	$ & $ e^{\pm i v_A k_{\parallel} t}				$ \\
\vphantom{$1 \over 2$} $ \textnormal{NADT-model}								$ & $ e^{\pm i v_A k_{\parallel} t} e^{- \beta t}	$ \\
\hline
\end{tabular}
\medskip
\label{tab1}
\end{table}
As in the most previous models we assume an exponential decorrelation of the magnetic fluctuations and, therefore
\be
\Gamma^{slab} (k_{\parallel},t) = e^{i \omega t - \gamma_c t}.
\ee
In a recent article (Lazarian \& Beresnyak 2006) is was shown that for slab modes
\be
t_c^{-1} = \gamma_c = v_A \sqrt{k_{\parallel} / L}
\label{Alexmodel}
\ee
and therefore
\be
\Gamma^{slab} (k_{\parallel},t) = e^{- v_A \sqrt{k_{\parallel} / L} \cdot t} \cdot e^{i \omega \cdot t}
\label{c2e8}
\ee
with the Alfv\'en speed $v_A$, the plasma box size $L$, and with the plasma wave dispersion relation of shear 
Alfv\'en waves 
\be
\omega = j v_A k_{\parallel}.
\label{c2e10}
\ee
The form of Eq. (\ref{Alexmodel}) can be justified by the following argument (for a more detailed 
explanation see Lazarian \& Beresnyak 2006): consider a wavepacket of Alfv\'en waves that
moves nearly parallel to the magnetic field with the dispersion of angles 
$\delta k_{\perp}/k_{\parallel} \sim \Theta_k$. The individual waves follow the local direction
of the magnetic field lines. As a result, the dispersion in angles of the wave packet cannot be
less than the dispersion of angles due to the ambient Alfv\'en turbulence, $\Theta_k > \Theta_{bk}$. 
The latter for the Goldreich-Sridhar (1995) model 
\be
k_{\parallel} \sim k_{\perp}^{2/3} L_A^{-1/3}, \quad \delta v \sim v_A (k_{\perp} L_A)^{1/3}
\label{eqal1}
\ee
is $\Theta_{bk} \sim \delta B_k / B_0 \sim (k_{\perp} L)^{-1/3}$. The modes with minimal $\Theta_k$ 
are the fastest growing ones. As we establish below (see Eq. (\ref{eqal2})) they are the least
damped. Therefore for our simplified treatment we shall limit our attention to the wavepackets
with resonant $k_{\parallel}^{-1} \sim r_p$ and $\Theta_{bk} \sim \Theta_{k}$. One can determine
the characteristic perpendicular wavenumber 
$k_{\perp} \sim \delta k_{\perp} \sim r_p^{-1} (r_p / L)^{1/4}$ of the most parallel modes that
are created by streaming CR's. The strong Alfv\'enic turbulence decorrelates the wavepacket
with $k_{\perp}$ on the time scale $v_{\perp} k_{\perp}$. Thus using the above expression for
$k_{\perp}$ and Eq. (\ref{eqal1}) we get
\be
t_c^{-1} = \gamma_c \sim -k_{\perp} v_{\perp} \sim -v_A k_{\perp}^{2/3} L^{-1/3}
\sim -v_A r_p^{-1/2} L^{-1/2}
\label{eqal2}
\ee
and thus Eq. (\ref{Alexmodel}), which, up to the ''-'' that we used to denote the damping
nature of the process, coincides with the damping rate obtained in Farmer \& Goldreich (2004)
and with the results of the numerical simulations shown in Lazarian \& Beresnyak (2006, Fig. 1 therein).

The parameter $j$ is used to track the wave direction ($j=+1$ is used for forward and $j=-1$ for backward to the ambient 
magnetic field propagating Alfv\'en waves). A lot of studies have addressed the direction of propagation of
Alfv\'enic turbulence, see for instance Bavassano (2003). In general one would expect that closer to the Sun
the most  waves should propagate forward and far away from the Sun the wave intensities should be equal for both
directions. In the current paper we are interested in turbulence parameters at 1 AU. Thus, we simply assume that
all waves propagate forward and we therefore set $j=+1$.

In the following we employ the dynamical correlation function defined by Eqs. (\ref{c2e8}) and 
(\ref{c2e10}) to determine the parallel mean free path within the quasilinear approach. 
\section{The quasilinear parallel mean free path}
In the current paper we employ quasilinear theory (QLT, Jokipii 1966) to calculate the parallel mean free path. QLT can be 
seen as a first order perturbation theory in the small parameter $\delta B / B_0$. Whereas it was shown previously
(e. g. Michalek \& Ostrowski 1996) that QLT is accurate even if $\delta B \approx B_0$ if we assume slab
geometry and a wave spectrum without dissipation range, it was realized by more recent test particle simulations
that for non-slab models and for steep wave spectra nonlinear effects are important and QLT is no longer accurate
(see e.g. Qin et al. 2006 and Shalchi 2007). In the current article we only investigate the slab model and we 
can therefore assume validity of QLT as it is shown in Shalchi et al. (2005) that the unphysical
contribution of large scale motions arising in QLT due to its inability to account for the conservation
of adiabatic invariant is small.
According to Jokipii (1966), Hasselmann \& Wibberenz (1968), Earl (1974), and Shalchi (2006), the parallel mean 
free path results from the pitch-angle-cosine average of the inverse pitch-angle Fokker-Planck coefficient 
$D_{\mu \mu }$ as
\be
\lambda_{\parallel}={3 v \over 8} \int_{-1}^{+1} d \mu \; {(1-\mu ^2)^2 \over D_{\mu \mu}(\mu )}.
\label{c4e1}
\ee
The pitch-angle-cosine $\mu$ is defined as $\mu=v_{\parallel}/v$. According to Teufel \& Schlickeiser (2002, Eq. 25) 
the pitch-angle Fokker-Planck coefficient can be written as
\bdm
D_{\mu \mu}^{i} (\mu) & = & { \Omega ^{2} (1-\mu^{2}) \over 2 B_{0} ^{2} } \sum _{n=-\infty} ^{+\infty} \int d^{3} k \; R_{n}^{i} (\vec{k})\nonumber\\
& & \left[ J_{n+1}^{2} \left( {k_{\perp} v_{\perp} \over \Omega} \right) P_{RR}^{i} (\vec{k}) \right. \nonumber\\
& + & \left. J_{n-1}^{2} \left( {k_{\perp} v_{\perp} \over \Omega} \right) P_{LL}^{i} (\vec{k}) \right. \nonumber\\
& - & J_{n+1} \left( {k_{\perp} v_{\perp} \over \Omega} \right) J_{n-1} \left( {k_{\perp} v_{\perp} \over \Omega} \right) \nonumber\\
& \times & \left. \left( P_{RL}^{i} (\vec{k}) e^{+2 i \Psi} + P_{LR}^{i} (\vec{k}) e^{-2 i \Psi} \right) \right]
\label{c3e3}
\edm
if we use helical coordinates
\bdm
\delta B_{L} & = & {1 \over \sqrt{2}} \left( \delta B_x + i \delta B_y \right), \nonumber\\
\delta B_{R} & = & {1 \over \sqrt{2}} \left( \delta B_x - i \delta B_y \right)
\label{c3e4}
\edm
and if we neglect electric fields \footnote{ 
Because of the high conductivity of cosmic plasmas, there are no large-scale electric fields 
$\left< \vec{E} \right> = \vec{E}_0 = 0$ and we thus have
$\vec{B} = B_0 \vec{e}_z + \delta \vec{B}, \vec{E} = \delta \vec{E}$ with the turbulent 
electric and  magnetic fields ($\delta \vec{E}$, $\delta \vec{B}$). The reason for using the model of 
purely magnetic fluctuations is that the electric fields are much smaller than the magnetic fields,
since we have $\delta E_i \sim {v_A \over c} \delta B_j \ll \delta B_j$, with the
Alfv\'en speed $v_A$ which is much smaller than the speed of light, $c$.}.
In Eq. (\ref{c3e3}) we used the resonance function
\be
R_{n}^{i} (\vec{k}) = Re \int_{0}^{\infty} d t \; e^{-i(k_{\parallel} v_{\parallel} + n \Omega) t} \cdot \Gamma^{i} (\vec{k},t)
\label{c3e6}
\ee
where the index $i$ stands for the different turbulence models. For pure slab geometry we have according to 
Teufel \& Schlickeiser (2002, Eq. 33)
\bdm
P_{RR} (\vec{k}) & = & P_{LL} (\vec{k}) = g^{slab} (k_{\parallel}) {\delta (k_{\perp}) \over k_{\perp}}, \nonumber\\
P_{RL} (\vec{k}) & = & P_{LR} (\vec{k}) = 0
\label{c3e7}
\edm
if we assume vanishing magnetic helicity. For the pitch-angle Fokker-Planck coefficient we then find
\bdm
D_{\mu\mu}^{slab} & = & {\pi \Omega^2 (1-\mu^2) \over B_0^2} \int_{-\infty}^{+\infty} d k_{\parallel} \; g^{slab} \left( k_{\parallel} \right) \nonumber\\
& \times & \sum_{n=\pm1} R_{n}^{slab} (k_{\parallel}).
\label{c3e8}
\edm
The resonance function for pure slab turbulence has the form
\be
R_{n}^{slab} (k_{\parallel}) = Re \int_{0}^{\infty} d t \; e^{-i(k_{\parallel} v_{\parallel} + n \Omega) t} \cdot \Gamma^{slab} (k_{\parallel}, t).
\label{c3e9}
\ee
With Eq. (\ref{c2e8}) for $\Gamma^{slab} (k_{\parallel}, t)$, the integral in Eq. (\ref{c3e9}) is elementary and we obtain
\be
R_{n}^{slab} = {v_A \sqrt{k_{\parallel} / L} \over v_A^2 k_{\parallel} / L + (k_{\parallel} v_{\parallel} + n \Omega - v_A k_{\parallel})^2}.
\label{c3e11}
\ee
With this Breit-Wigner-type resonance function the slab Fokker-Planck coefficient can be written as
\bdm
D_{\mu\mu}^{slab} & = & {2 \pi \Omega^2 (1-\mu^2) \over B_0^2} \int_{0}^{\infty} d k_{\parallel} \; g^{slab} (k_{\parallel}) \nonumber\\
& \times & \sum_{n=\pm1} {v_A \sqrt{k_{\parallel} / L} \over v_A^2 k_{\parallel} / L + (k_{\parallel} v_{\parallel} + n \Omega - v_A k_{\parallel})^2}.
\label{c3e12}
\edm
With the integral transformation $x=l_{slab} k_{\parallel}$ and with the parameters $R=R_L / l_{slab}=v / (\Omega l_{slab})$
and $\epsilon = v_A / v$ we obtain
\bdm
D_{\mu\mu}^{slab} & = & {2 \pi (1-\mu^2) \over B_0^2 l_{slab}} \int_{0}^{\infty} d x \; g^{slab} \left( k_{\parallel} = {x \over l_{slab}} \right) \nonumber\\
& \times & \sum_{n = \pm 1} {v_A \sqrt{k_{\parallel} / L} \over (v_A \sqrt{k_{\parallel} / L} / \Omega)^2 + \left[ x R (\mu - \epsilon) + n \right]^2}.
\label{c3e13}
\edm
The slab spectrum of Eq. (\ref{c2e5}) can be written as 
\be
g^{slab} (x) = {C(\nu) \over 2 \pi} l_{slab} \delta B_{slab}^2 \; h^{slab}(x,x_{min},\nu,\xi,p)
\label{c3e14}
\ee
with
\bdm
& & h^{slab}(x,x_{min},\nu,\xi,p) \nonumber\\
& = & \left\{
\begin{array}{ccc}
0 & \textnormal{for} & 0 \leq x < x_{min} \\
(1+x^2)^{-\nu} & \textnormal{for} & x_{min} \leq x < \xi \\
(1+\xi^2)^{-\nu} ({\xi \over x})^{p} & \textnormal{for} & \xi \leq x
\end{array}
\right.
\label{c3e15}
\edm
where we used $\xi=l_{slab} k_{d}$. Then we find for the dimensionless Fokker-Planck coefficient 
$\tilde{D}_{\mu\mu}^{slab} = D_{\mu\mu}^{slab} l_{slab} / v$
\bdm
& & \tilde{D}_{\mu\mu}^{slab} = {C(\nu) (1-\mu^2) \over R} {\delta B_{slab}^2 \over B_0^2} \nonumber\\
& \times & \int_{0}^{\infty} d x \; h^{slab} (x,x_{min},\nu,\xi,p) \nonumber\\
& \times & \sum_{n=\pm1} {v_A \sqrt{x / (l_{slab} L)} / \Omega \over (v_A \sqrt{x / (l_{slab} L)} / \Omega)^2 + \left[ x R (\mu - \epsilon) + n \right]^2}.
\label{c3e16old}
\edm
By using 
\be
{v_A \over \Omega} \sqrt{x \over l_{slab} L} = {v_A \over v} {v \over \Omega l_{slab}} \sqrt{ x {l_{slab} \over L}} = \epsilon R \sqrt{x \eta}
\label{c3e17}
\ee
we finally find
\bdm
\tilde{D}_{\mu\mu}^{slab} & = & {C(\nu) (1-\mu^2) \over R} {\delta B_{slab}^2 \over B_0^2} \nonumber\\
& \times & \int_{0}^{\infty} d x \; h^{slab} (x,x_{min},\nu,\xi,p) \nonumber\\
& \times & \sum_{n=\pm1} {\epsilon R \sqrt{x \eta} \over (\epsilon R \sqrt{x \eta})^2 + \left[ x R (\mu - \epsilon) + n \right]^2}.
\label{c3e16}
\edm
In Eqs. (\ref{c3e14}) - (\ref{c3e16}) we used $\rho=l_{slab}/l_{2D}$ and $x_{min} \equiv \eta = l_{slab}/L$. A numerical investigation 
of the integral of Eq. (\ref{c3e16}) is presented in the next section.
\section{Numerical results for pitch-angle and parallel diffusion}
In this section we evaluate the formulas for pitch-angle diffusion (Eq. (\ref{c3e16})) and for the mean free 
path (Eq. (\ref{c4e1})) derived in the last sections numerically for the parameter-set of Table \ref{tab2} 
which should be appropriate for interplanetary conditions at 1 AU heliocentric distance. All formulas depend 
on the parameter $\epsilon$ which can be expressed as (see Shalchi et al. 2006)
\be
\epsilon = {v_A \over v} = {v_A \over c} {\sqrt{R_0^2 + R^2} \over R}
\label{c6e1}
\ee
with
\bdm
R_0 = {1 \over l_{slab} \cdot B_0} \cdot
\left\{
\begin{array}{ccc}
0.511 MV & \textnormal{for} & e^{-} \\
938 MV & \textnormal{for} & p^{+}
\end{array}
\right.
\label{c6e2}
\edm
For the heliospheric parameters considered in the current paper we have for electrons 
$R_0 (e^{-}) \approx 9.2 \cdot 10^{-5}$ and protons $R_0 (p^{+}) \approx 0.169$. 

In order to determine the transport coefficients of the isotropic part of the particle distribution 
function, e.g. the parallel mean free path, we must restrict our calculations to $\epsilon = v_A / v \ll 1$ 
(see Schlickeiser 2002 for a detailed explanation). Thus, we can only consider rigidities which satisfy 
the following condition:
\bdm
R \gg {R_0 \over \sqrt{\left( c / v_A \right)^2 - 1}} \approx R_0 {v_A \over c}
\label{c6e2a}
\edm
For $v_A=33.5 \; km/s$ we find for electrons the restriction $R (e^{-}) \gg 10^{-8}$ and for protons $R (p^{+}) \gg2 \cdot 10^{-5}$.
\begin{table}
\caption{Parameters used for our numerical calculations. The values should be appropriate for the solar wind
at 1 AU heliocentric distance. If a parameter is different from the values below we note this separately in 
the corresponding figures and discussions. These values correspond to the values used in Shalchi et al. (2006).}
\begin{tabular}{|l|l|l|}\hline
\vphantom{$1 \over 2$} $ \textnormal{Parameter}						$ & $ \textnormal{Symbol}		$ & $ \textnormal{Value}		$ \\ 
\hline\hline
\vphantom{$1 \over 2$} $ \textnormal{Inertial range spectral index}			$ & $ 2 \nu					$ & $ 5/3					$ \\
\vphantom{$1 \over 2$} $ \textnormal{Dissipation range spectral index}		$ & $ p						$ & $ 3 						$ \\
\vphantom{$1 \over 2$} $ \textnormal{Alfv\'en speed}						$ & $ v_A					$ & $ 33.5 \; km/s				$ \\
\vphantom{$1 \over 2$} $ \textnormal{Slab bendover scale}				$ & $ l_{slab}				$ & $ 0.030 \; AU				$ \\
\vphantom{$1 \over 2$} $ \textnormal{Slab dissipation wavenumber}		$ & $ k_{slab}				$ & $ 3 \cdot 10^{6} \; (AU)^{-1}	$ \\
\vphantom{$1 \over 2$} $ \textnormal{Mean field}						$ & $ B_0					$ & $ 4.12 \; nT				$ \\
\vphantom{$1 \over 2$} $ \textnormal{Turbulence strength}				$ & $ \delta B_{slab} / B_0		$ & $ 1						$ \\
\hline
\end{tabular}
\medskip
\label{tab2}
\end{table}
\subsection{The pitch-angle Fokker-Planck coefficient $D_{\mu\mu}$}
\begin{figure}
\begin{center}
\epsfig{file=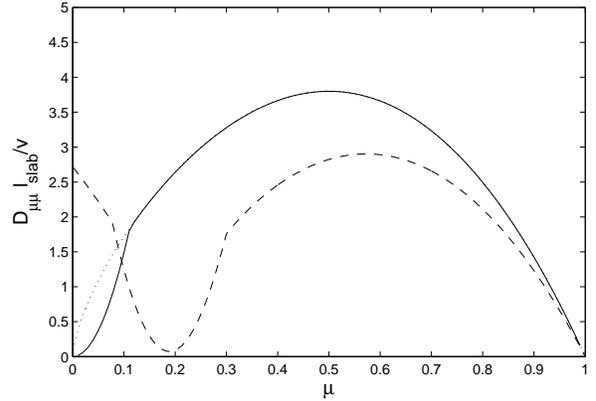, width=250pt}
\end{center}
\caption{The pitch-angle Fokker-Planck coefficient for protons (dashed line) and electrons (solid line).
Also the magnetostatic dissipationless model is shown (dotted line). Visualized is a linear plot of the 
dimensionless pitch-angle Fokker-Planck coefficient $\tilde{D}_{\mu\mu}$ as a function of the pitch-angle 
cosine $\mu$. All results are for pure slab geometry and for small rigidities $R=10^{-4}$.}
\label{nadtf23}
\end{figure}

Fig. \ref{nadtf23} shows the (dimensionless) pitch-angle Fokker-Planck coefficient $\tilde{D}_{\mu\mu}=D_{\mu\mu} l_{slab}/v$
as a function of the pitch-angle-cosine $\mu$. For the rigidity we assumed $R=10^{-4}$. In this case the parameter
$\epsilon$ (see Eq. (\ref{c6e1})) is approximately $\epsilon \approx 0.2$. It seems that the minimum of the pitch-angle Fokker-Planck coefficient for protons can be found at $\mu \approx \epsilon$.
In general the pitch-angle Fokker-Planck coefficient is no longer equal to zero at $90^o$ ($\mu=0$) as in the magnetostatic model,
so that we no longer obtain an infinitely large parallel mean free path as in magnetostatic models. It should be noted, however, that 
QLT itself is questionable close to $90^o$. By considering Fig. \ref{nadtf23} we find that at least for protons 
pitch-angle scattering close to $90^o$ is very strong due to the dynamical effects. Therefore one could assume that nonlinear 
effects which also lead to nonvanishing pitch-angle scattering at $90^o$ could be neglected.
\subsection{The parallel mean free paths $\lambda_{\parallel}$}
\begin{figure}
\begin{center}
\epsfig{file=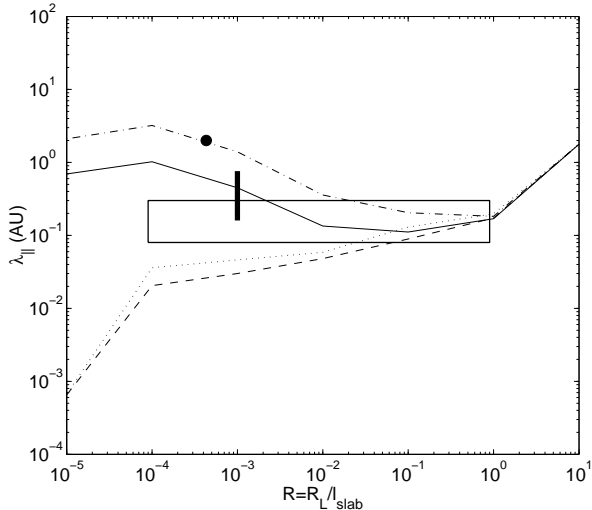, width=250pt}
\end{center}
\caption{The quasilinear parallel mean free path $\lambda_{\parallel}$ for $x_{min}=\eta=0.1$  for electrons (solid line) and 
protons (dashed line), and for $x_{min}=\eta=0.01$ for electrons (dash-dotted line) and protons (dotted line) in comparison 
with different solar wind observations: Palmer consensus (Palmer 1982, box), Ulysses observations (Gloeckler et al. 1995, dot), 
AMPTE spacecraft observations (M\"obius et al. 1998, vertical line).}
\label{nadtf4}
\end{figure}

Here we present theoretical results for the parallel mean free path. We compare our results with the Palmer consensus 
(Palmer 1982) and pickup ion observations (Gloeckler et al. 1995, M\"obius et al. 1998). In Fig. \ref{nadtf4}
we show the parallel mean free path in comparison with observations. We computed the parallel mean free paths for
two different values of the smallest wave number $x_{min}$, namely for $x_{min}=0.1$ and $x_{min}=0.01$.
%
%
%\begin{figure}
%\begin{center}
%\epsfig{file=Alex4.eps, width=250pt}
%\end{center}
%\caption{The parallel mean free path $\lambda_{\parallel}$ for $x_{min}=\eta=0.01$. Shown are QLT results for electrons 
%(solid line) and protons (dashed line) in comparison with different observations: Palmer consensus (Palmer 1982, box), 
%Ulysses observations (Gloeckler et al. 1995, dot), AMPTE spacecraft observations (M\"obius et al. 1998, vertical line).}
%\label{nadtf5}
%\end{figure}
%
%
In addition to the Palmer (1982) results we compare our results also with pickup ion observations:
\begin{enumerate}
\item  Gloeckler et al. (1995) concluded from Ulysses observations that the parallel mean free paths of pickup protons 
is 2 AU at 2.4 MV rigidity (they stated conservatively that $\lambda_{\parallel}$ is of order 1 AU but actually they obtained 
the best fit for 2 AU).  It should be noted that this observation was at high heliographic latitudes, and at a heliocentric distance 
of 2.34 AU; these differences should be remembered when comparing with observations at Earth orbit.
\item M\"obius et al. (1998) concluded from AMPTE (Active Magnetospheric Particle Tracer Explorers) spacecraft observations 
that the parallel mean free paths of pickup helium ranges from 0.16 to 0.76 AU at 5.6 MV rigidity in the data they analyzed.
\end{enumerate}
Both results are also illustrated in Fig. \ref{nadtf4}. As shown we can reproduce the observations theoretically.
%The M\"obius et al. (1998) observations are close to our 
%theoretical results but the Gloeckler et al. (1995) measurements are larger than our theoretical predictions. We expect 
%that the reason for this discrepancy is that the Gloeckler et al. (1995) observation was at high heliographic latitudes, and at a 
%heliocentric distance of 2.34 AU. For such conditions the turbulence parameters are expected to be quite different from the 
%values we used for our theoretical calculations (see Table \ref{tab2}).

\begin{figure}
\begin{center}
\epsfig{file=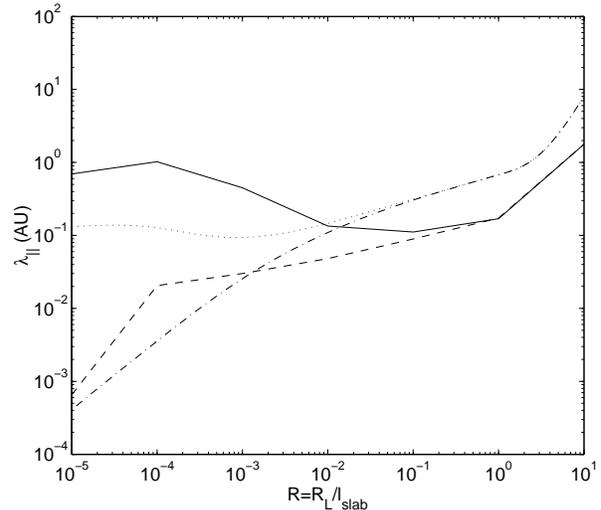, width=250pt}
\end{center}
\caption{The parallel mean free path $\lambda_{\parallel}$ for $x_{min}=\eta=0.1$. Shown are QLT results for electrons 
(solid line) and protons (dashed line) in comparison with the theoretical results obtained by Shalchi et al. (2006) by
applying the NADT-model in combination with the slab/2D model for the turbulence geometry (results for electron
are represented by the dotted line, and the results for protons are represented by the dash-dotted line).}
\label{nadtf6}
\end{figure}

In Fig. \ref{nadtf6} we compare our new theoretical results with the results obtained by Shalchi et al. (2006) by
applying the NADT-model in combination with the slab/2D model for the turbulence geometry. 
As shown, for small rigidities, where dissipation effects are important, we obtain a much larger parallel mean 
free path for electrons. For medium rigidities where the charged particles interact resonantly with the inertial 
range of the spectrum, the new parallel mean free path is about a factor 5 smaller\footnote{That fact that the 
parallel mean free path is much smaller for pure slab geometry (in comparison to the 20\% slab / 80\% 2D result
of Shalchi et al. 2006) and medium rigidities, can be seen as a trivial result, since within QLT we find no gyro-resonant 
scattering due to 2D modes for magnetostatic turbulence.} since we assumed pure slab fluctuations (in comparison to 
the slab/2D\footnote{The slab/2D model used in Shalchi et al. 2006 assumes a superposition of pure slab and pure 2D 
modes. It is well known that this model can only be approximatelly correct, since real turbulence has a certain distribution 
of wave vectors in the $\vec{k}-$space.} model used in Shalchi et al. 2006). It seems that the improved model for the 
slab correlation function provides a much larger electron parallel mean free path for low energetic particles.
\section{Summary and Conclusion}
The theoretical explanation of measured parallel mean free paths in the solar system is a fundamental problem of
space science. In a recent article (Shalchi et al. 2006) it has been demonstrated the these observations
can indeed be reproduced theoretically. By using recent results of turbulence theory (Lazarian \& Beresnyak 2006)
we further improved the dynamical correlation function which is a key input in transport theory considerations.
It is demonstrated in this article that the improved slab correlation time scale (see Eq. (\ref{Alexmodel})) leads to a 
much larger parallel mean free path (see Fig. \ref{nadtf4}). This effect is important since it was argued 
in several previous articles that the theoretical parallel mean free path is too small (Palmer 1982, Bieber et al. 1994)
in comparison with solar wind observations. 

Another problem of cosmic ray scattering theory is the importance of nonlinear effects. Whereas we have
applied QLT in the current article it was argued in other papers (e.g. Shalchi et al. 2004) that nonlinear effects
are important for parallel diffusion. However, these nonlinear effects are directly related to the interaction
between charged particles and 2D modes. These 2D modes were neglected since we assumed pure slab
fluctuations. Therefore, QLT can be applied and the results presented in this article should be valid.
For non-slab models, where 2D modes are present, however, the applicability of QLT is questionable.
It has to be subject of future work to explore the validity of QLT for realistic turbulence models
such as dynamical turbulence models in non-slab geometry.
\section*{Acknowledgments}
This research was supported by Deutsche Forschungsgemeinschaft (DFG) under the Emmy-Noether program
(grant SH 93/3-1). This work was also supported by the NASA grant X5166204101, the 
NSF grant ATM-0648699, and the NSF Center for Magnetic Self Organization in Laboratory and 
Astrophysical Plasmas.
This work is the result of a collaboration between the University of Bochum, Theoretische Physik IV and
the University of Wisconsin-Madison, Department of Astronomy.
{}

\end{document}